\documentclass[12pt,reqno]{article}
\usepackage{latexsym,amsmath,amsthm,amsfonts,amssymb,amscd}
\usepackage[dvips]{graphicx,psfrag}

\title{\bf Probability as typicality}
\author{S\'ergio B. Volchan}

\begin{document}

\date{26 October  2006}
\maketitle

\vskip .5cm
\begin{flushright}
\end{flushright}

\begin{abstract}

The concept of typicality refers to  properties holding for
the ``vast majority'' of cases and is a fundamental idea of
the qualitative approach to dynamical problems. We argue that 
measure-theoretical typicality would be the adequate viewpoint 
of the role of probability in classical statistical mechanics, 
particularly  in understanding the  micro to macroscopic change of 
levels of  description.

\end{abstract}



\bigskip

\noindent{\bf 1. Introduction}

 The year 2006 marked the $100$th anniversary of  Ludwig Boltzmann's  death. He is 
justly celebrated as one of the greatest theoretical physicists of the XIXth century 
and a founding father of statistical  mechanics (for a perspective, see Cercignani, 
1998 and Uffink, 2004). Besides, his work  influenced (directly or indirectly) the 
development of important fields of pure and applied mathematics: foundations of 
probability theory, the theory of stochastic processes (Brownian motion), ergodic theory
and functional analysis (integral-differential equations).

He also had a significant impact on the philosophy of science. Firstly, 
he contributed to the unity of physics by attempting to reconcile the 
atomistic-mechanical view of microscopic dynamics with the 
macroscopic world of thermodynamics in the context of kinetic gas theory.
Secondly, in  that endeavor he had a clear view that progress in physics 
involves  hypothesizing unobservable (but not inscrutable) 
``simple'' material entities in order to explain
the behavior of ``complex'' systems.~\footnote{In stark contrast to the somewhat 
barren positivist stance of the ``energeticists'' (according to which energy was 
a kind  of ``primordial'' entity) for which  unobservables had no
place in physics. Curiously,  they regarded thermodynamics as their ideal, a theory which
is full of unobservables (as in any  deep physical theory, Bunge, 1967),
entropy being one of the most intangible.}
Thence his role in the ``battle over the reality of the  atom'' (Wick, 1995), 
which was the cornerstone of that ``unification'' 
program and which served as a prelude to the future unraveling 
of the structure of matter (see Brush, 1994).   Thirdly, and closer to the focus of 
this article, in facing the sharp criticisms  (the
so-called ``reversibility paradoxes'') against his program, he brought
(with the contribution of Gibbs)  probabilistic arguments to the center of stage 
in physics. 

However, the use probabilistic concepts in kinetic theory~\footnote{Interestingly, 
the use of statistics in kinetic gas theory was inspired by the astronomer and
demographer Adolphe Qu\^etelet,  see Torretti, (1999), p. 180 and von Plato (1998), p. 73.}
has generated a lot of controversy and  misunderstanding, even after so many  
discussions and clarifications. 
For, how come that starting from  Newton's equations  of motion for discrete
particles  one suddenly conjures up a continuous probability density in 
phase-space?~\footnote{It seems that the idea of studying a mechanical systems by
means of a probability density on the set of initial data was pioneered by 
Poincar\'e in his ``method of arbitrary functions'', see von Plato, 1983.}
What exactly is the role and status of probability in classical statistical mechanics? Is it 
a concept totally alien to mechanics?

The difficulties Boltzmann had to face were not only conceptual but also technical  
due to the lack of adequate  mathematical tools  necessary to  tackle 
(and even formulate) such hard questions in a clear,
rigorous and meaningful way. For example, one must bear in mind that a 
probability {\em theory}  proper was  not yet available and there was (and still is!) 
some confusion  regarding its status, as many  people viewed it as a peculiar blend of physics 
and  mathematics.~\footnote{According to  Kac (1949), Poincar\'e used to joke about the 
confusing status of the central limit theorem, saying that ``there must be something mysterious
about the normal law since  mathematicians think it is a law of nature whereas physicists
are convinced it is a mathematical theorem''  (p. 52) (the physicists were 
right here!). Maybe a trace of this confusion remains as many of the {\em theorems} of
probability theory are labeled  as ``laws'',  as in the ``law of large numbers'', 
``the normal law'', ``Kolmogorov 0-1 law'', etc. Incidentally, in 2006 a Fields Medal (the
Nobel prize of mathematics) was  awarded for the first time ever to a probabilist.} 
As von Plato's (1991) pointed out,  Boltzmann ``was a XIXth-century theoretical physicist, 
not a XXth-century mathematician. Even so, he has been judged and interpreted according to 
mathematical concepts and standards that were not his. The mathematics of probability of 
the previous century being what it was, he was still able to achieve his end by reasoning 
that later developments confirmed'' (p. 87).

That the problems of kinetic theory were deemed  highly challenging (in fact, they are
still open today) and important can be gauged by their inclusion  as part of the 6th  problem 
in  David  Hilbert's famous list. It deals  with the axiomatization of physical 
theories   and, 
in particular,  probability theory.~\footnote{Note that Hilbert considered  probability 
(like geometry) as part of physics, not an uncommon view at the time.}
In his words (quoted in Corry, 1997, p. 121):
\begin{quote}
 The investigations on the foundations of geometry suggest the 
problem: To treat in the same manner, by means of axioms, those physical
sciences in which mathematics plays an important part; in the first rank
are the theory of probabilities and mechanics.
 
As to axioms of the theory of probability, it seems to me desirable
that their logical investigation should be accompanied by a rigorous and
satisfactory development of the method of mean values in mathematical
physics and in particular in the kinetic theory of gases.

\end{quote}

In this paper we argue that {\em typicality} would be the
adequate viewpoint of the role of probability in classical statistical mechanics.~\footnote{The notion
of typicality also plays a crucial role in the revamped version of Bohmian  quantum mechanics, see for
instance D\"urr (2001).}
Instead of indicating the presence of a random ingredient in the system, it is taken 
as a yardstick in probing the relative {\em size} of some sets of micro-states of interest
(in particular, initial conditions) in the geometrical arena of phase-space. As discussed in 
section 2, this view is at the very heart of  modern axiomatic probability theory, 
which is based on Borel and Lebesgue's measure theory. Now, as the initial  data are an 
integral part of any mechanical  system, the use of   probability to size up such data makes 
it less foreign to mechanics.

 Generally speaking, a set is typical if it contains an  ``overwhelming majority'' of points 
in some specified sense.  In classical statistical mechanics there is a ``natural'' sense: namely
sets of full phase-space volume. That is, one considers the Lebesgue-measure~\footnote{More generally, 
an  absolutely continuous measure with respect to Lebesgue-measure.} on phase-space, which is invariant 
(by Liouville's theorem) which, when ``cut to the energy surface'', can be normalized to a probability 
measure; then sets of volume close to one are considered typical. We suggest that the focus 
on such (measure-theoretical)  typical micro-states leading to the system's macroscopic  
behavior (in an appropriate limit) underlines the role of probability  in bridging the micro-macro levels 
of description, which is a basic aim of statistical mechanics. Besides, as discussed  in section 3,  
probability as typicality has a long history of success, from celestial  mechanics to the modern theory of 
dynamical systems. 

We also think that, from  this perspective, the use of probabilistic reasoning in classical
statistical mechanics is more understandable. It could be conceived as part of the {\em zeitgeist} of 
the last decades of the XIXth when a revolutionary trend from quantitative to qualitative methods was 
taking place, 
pioneered by Poincar\'e, more or less at the same  time as Boltzmann's work in kinetic 
theory~\footnote{This would deserve a more thorough  historical investigation. It is fascinating
to wonder what these two quite different characters would had to say to each other as they most
probably met at the 1904 St. Louis world fair (Cercignani, 1998, p. 145).}
(a field in  which, by the way, Poincar\'e had a keen interest, his paper on the
subject dating from 1894, see von Plato, 1991, p. 83).

Finally, in section 4 we discuss the use of probability as typicality in a remarkable
achievement in non-equilibrium statistical mechanics,  namely, Lanford's theorem on the 
validity problem for Boltzmann's equation.
\medskip

\noindent{\bf 2. Probability and Measure Theory}

A. N. Kolmogorov solved  part of Hilbert's 6th problem in 1933 by his
axiomatization of probability theory in his {\em Grundbegriffe} (Kolmogorov, 1957).
The fundamental insight was that probability theory, going beyond  its ``elementary'' 
part which  deals with discrete sample spaces and reduces essentially to combinatorics, could be
seen as a  branch of the  newly created  {\em measure theory} of Lebesgue.

Although  measure theory developed from internal problems in mathematical analysis, 
linked to the need to generalize  Riemann's integral in the context of  Fourier's 
series (Kahane, 2001,  Hoare and Lord, 2002), it also had ancient geometric roots. As made clear in  Lebesgue's 1902 
doctoral  dissertation, titled {\em Int\'egrale, longueur, aire}, measure theory is conceived as 
a (very abstract) generalization of basic geometrical notions of size: length, area and
volume (Choquet, 2004).

Recall that in  modern mathematical language a {\em measure space} is a 
triple $(\Omega, {\mathcal F},\mu)$  where: $\Omega$ is an arbitrary set 
(usually equipped with some natural topology);
$\mathcal F$ is a collection of subsets of $\Omega$, called {\em measurable subsets}, 
carrying the structure of a $\sigma$-algebra (i.e., it is closed under denumerable
set-theoretical operations); and $\mu$ is a non-negative  countably 
additive set function on $\mathcal F$.~\footnote{A pair $(\Omega, \mathcal F)$ is 
called a {\em measurable space}, meaning that it can carry different measures.} 
Kolmogorov noticed that probability theory could be perfectly
couched  in this framework. A  {\em probability  space} is defined as a 
triple  $(\Omega, \mathcal F, \mathbf P)$, where $\Omega$ is the ``sample space'',
$\mathcal F$ the collection of  ``events'' and $\mathbf P$ is  a finite measure 
normalized to one, ${\mathbf P}(\Omega) =1$. If $A$ is an ``event''
(a measurable set)~\footnote{Notice that, as   $\mathcal F$ is usually strictly 
smaller than the set of all subsets of $\Omega$, some subsets of the sample space 
may have no probability at all.}, then  ${\mathbf P}(A)$ is its probability ``of occurrence''. 
Also, ``random variables'' are identified to measurable functions and expectations to Lebesgue-integrals with
respect to the given probability measure.~\footnote{Of course, Kolmogorov's legacy in probability theory is 
much wider: he clarified the concept of independence (there is a joke saying that probability theory 
is ``just'' measure theory plus the concept of independence in the same way that complex analysis 
as ``just'' analysis in two-variables plus $\sqrt -1$) and conditioning; established a host of
now-classical limit theorems for sums of independent random variables; and made seminal 
contributions to the theory of continuous-time stochastic 
processes (see Mazliak, Chaumont and Yor, 2004).} 

Kolmogorov's axiomatization  (which today reached almost universal acceptance) has an enormous 
significance: not only it gave the seal of maturity and mathematical respectability to the discipline, 
which has been expanding relentlessly  since then, as it  greatly clarified its nature. Most importantly, 
it became clear, once and for all, that probability theory is a branch of pure mathematics, like group theory, 
geometry, linear algebra,  etc. Hence it has many models in the set-theoretic sense so that the expression
 ``taken at random'' has different meanings depending on the specified
probability space. This helped dissolving many paradoxes that plagued
probability theory, like Bertrand's, which were linked to a careless use of that 
phrase. Freed from any previous commitment to
an ``interpretation'' (be it frequentist, subjectivist, propensity, etc) the theory 
could be developed autonomously. Moreover, once 
its nature is so elucidated one can  examine and criticize any proposed interpretation or 
application of probability theory to the real-world.

For the better or worse,  probability theory has kept the old jargon of its pre-axiomatized 
era. Though it is debatable whether  such notions as ``sample'', ``event'', ``occurrence'',
``trials'', ``favorable event'', ``experiment'', etc may or not have some heuristic or 
pedagogic value, the fact is that they are not, strictly speaking, part of probability 
theory. In particular, there is nothing intrinsically random about  
``random variables'': they are just real-valued measurable functions, those  
one expects to find in real analysis (for instance, the continuous functions).

The old phraseology  has to be used with great care, particularly  in applications.
For instance,  coin-tossing is usually taken as the epitome of a random phenomenon.
However, {\em real} coin-tossing is a purely mechanical process that should in principle
be modeled using rigid-body Newtonian dynamics plus the initial conditions; and its relation
to randomness and unpredictability is a  non-trivial matter (see Diaconis, Holmes and
Montgomery, 2005).

\bigskip

\noindent{\bf 3. Negligible sets}

As a measure on a set can be viewed as giving the size 
of some of its subsets, so a probability measure can
be conceived as giving their relative sizes. Moreover, a measure allows us to
take some sets as being ``small'', ``exceptional'',  ``atypical'' or 
``negligible'' and hence ignored  in some specific contexts.

For the record, we note that there are at least two other common notions of size 
(and hence, of typicality) in mathematics: {\em cardinality}, giving the number of 
points (or power) of a set and a topological notion, called {\em genericity}, 
particularly  useful in dynamical systems theory. We won't discuss them here as 
our focus is on the {\em measure-theoretical} notion of 
typicality.~\footnote{The relationships between these three notions are quite complex,
see Oxtoby, 1971.} Roughly speaking, a property $\mathcal P$ on a measure space is {\em typical} if 
the set $\wp^c=\Omega-\wp$ of its {\em exceptions} has ``small'' 
measure (in particular, $\wp^c$ has  to be a measurable set)  that is, 
$\mu (\wp^c)\leq \epsilon$ for some ``tolerance''  $0\leq \epsilon \ll 1$, where 
$\wp =\{\omega \in \Omega: {\mathcal P}(\omega)\}$. It is important to
bear in mind that, as Goldstein (2001) observed, typicality ``plays solely the role of
informing us when a set $E$ of exceptions is sufficiently small that we may in effect ignore it''
(p. 53), so one could conceive of some weaker notion of typicality, without the additional 
measure-theoretical structure.~\footnote{A possibility that comes to mind is outer measure, which is 
only sub-additive and defined on all subsets of $\Omega$.}

An important and natural example are sets of measure zero (also call null sets), corresponding to
$\epsilon =0$.~\footnote{This is the strongest measure-theoretical notion of typicality and
historically the concept of null set predates measure theory proper, see von Plato (1983), p.44.}
In measure theory it is known that changing a measurable function on 
a null set (which can be uncountable)  won't affect the Lebesgue integral of that function.
This suggests the  notion of a property holding  {\em almost-everywhere}, meaning that  it holds 
outside a set of measure zero. In the  analogous situation in probability theory one
says it holds {\em almost-surely} or {\em with probability one}. 

Thus,  in measure/probability theory  a property is said to hold even if it has infinitely (even uncountably) 
many exceptions, as  long as these form a set of small measure/probability.  For instance,  in 1909 Borel 
obtained the first proof of a strong law of large numbers by showing that Lebesgue almost every real 
number in $[0,1]$ is normal to base 10, meaning that any block of $k$ digits appears with asymptotic 
frequency $10^{-k}$ in its decimal expansion (and the same holds for any base). So, though there are
uncountably many non-normal numbers, as seen through the lenses of Lebesgue measure (so to speak), ``all''
real numbers are normal.  However, as remarked by Kac (1949), ``as is often the case, it is much easier 
to prove that an overwhelming majority of objects possess a certain property than to {\em exhibit} 
even one such object'' (p. 18) and to this day no one knows whether such fundamental constants 
as $\pi$, $e$, $\sqrt 2$ are  normal (to any base!).~\footnote{By the way, normality was initially taken to 
be a reasonable definition of randomness for sequences of digits. However, it  was abandoned as it was 
proven by Champernowne in 1933 that the number $0.12345678910111213...$ (the concatenation of the positive 
integers) is normal to base 10. For a discussion of randomness for sequences, see Volchan, 2002.} We stress 
though, that the whole point of typicality  arguments is that, for certain purposes, one would not need such 
detailed  information.

The relation of typicality to probability is subtle, as the following example illustrates. Consider the set of
all  binary strings of say, 200 bits. For example, one could think of it as the sample space associated
to the tossing 200 ideal fair coins (heads=1, tail=0). Now, in the overwhelming majority of strings the
total number of heads is between 50 and 150 (i.e., within 50 of the mean 100) and we may declare
such strings as  typical. Of course, the string 11...11, consisting of 200 heads, is not typical. However, 
it has the very same (and very small)  probability, namely $2^{-200}$, as each and every one of the $2^{200}$ 
possible strings.~\footnote{This is a version of the so-called ``paradox of randomness'', see Volchan, 2002.}

It seems that  the  first  typicality-like arguments appeared in the field of
celestial mechanics, in connection with the classical problem  of the stability of the
solar system. This came about after the rather slow  realization that one could not
expect to  ``explicitly'' solve most differential equations and, moreover, that
such a solution might be uninformative. A clear  illustration of this  state of affairs is the 
Newtonian gravitational  three-body problem. Though this is not an integrable system 
(in the precise  sense of Hamiltonian mechanics) it has an ``explicit'' or ``analytic'' solution, 
which was  found by Sundman in 1909 (see Henkel, 2001). He obtained a convergent series 
solution, valid  for all times,~\footnote{And which fulfills precisely the requirements stated in 
the celebrated king Oscar II prize problem, see Barrow-Green, 1997.} but whose rate of convergence is 
so slow~\footnote{It is estimated that
$10^{8,000,000}$ terms of the series would be necessary to reach the standard of accuracy in modern
ephemeris calculations, see Goroff, 1992, p. I25} as to render it virtually useless to 
extract interesting information about the long-time behavior of the system.

In that long historical trend in mathematical-physics,  which eventually led to the
switch from quantitative to qualitative methods pioneered  by
Poincar\'e, Lyapunov and others (see Laskar, 1992 and Chenciner, 1999), the focus changes from
a detailed analysis of {\em individual} solutions of a given system to the study of whole {\em families} of 
them (and of families of systems).  As the 1994  Fields medalist J.-C. Yoccoz puts it: 
``Broadly speaking, the goal of the theory of dynamical systems is, as it should be, to understand most of
the dynamics of most systems.'' (Yoccoz, 1995, p.247). Note that as Sundman's example shows,  resorting to 
qualitative methods (including statistics) is not necessarily linked to  the huge number of  equations one is 
dealing with nor to ignorance 
(or imprecision) of initial conditions, notwithstanding  a common claim in statistical 
mechanics   textbooks.
                 
One of the earliest examples of the qualitative approach to dynamical problems is the famous 
{\em  Poincar\'e's recurrence theorem}  which Poincar\'e himself called 
``stabilit\'e a la Poisson''  (and is also known as ``Poincar\'e-reversibility''~\footnote{Which is
on of several concepts of reversibility. A nice discussion of the subject can be found in Illner 
and Neunzert, 1987.}). This remarkably simple 
result, which appeared in his memoir on the three-body  problem (1890), is a forerunner of 
the ergodic  theorems and  is one of the few global results in dynamical systems theory.

Its  measure-theoretical version  (not the original one, as measure theory was not yet available)
says the  following. Consider a flow ${\bf T}_{t}$  (e.g., associated to the solutions of a differential equation) 
on a set  $\Omega$ and $\mu$ a  finite {\em invariant} 
measure. Then, almost  all points of $\Omega$ are recurrent, that is, the orbit  of every initial condition  
$x$ outside a set of $\mu$-measure zero will eventually come arbitrarily close to  
$x$ (such orbits were called ``stable'').

Now, by normalizing $\mu$ to a probability measure the theorem can be rephrased thus:
the flow is recurrent with probability one or that an initial data ``taken at random'' is 
recurrent. According to von Plato (1991), in  this result  ``for the first time, a property 
is ascribed to mechanical systems with probability one. Exceptions to the problem are not impossible 
but have probability zero'' (p. 83).~\footnote{Incidentally,  from discussions of earlier work 
of Gyld\'en (1888) on the related  problem of planetary mean motion  came out the
first clearly articulated methodological principle linked to the negligibility 
of null sets, due to Felix Bernstein (1912). He  called it  ``the axiom of
the limited arithmetizability of observations''  according to which (quoted in von Plato, 1998, p.63):
\begin{quote}
 When one relates the values of an experimentally measured quantity to the
scale of all the reals, one can exclude from the latter in advance any set of
measure zero. One should expect only such consequences of the observed events which
are maintained when the observed value is represented by another one within the
interval of observation.

\end{quote}} 
However, as there is no intrinsic randomness in the dynamics nor 
in the initial data, what the theorem really seems to convey  is that the property of recurrence is 
typical with  respect to the measure $\mu$. In other words, it holds for ``the vast majority'' 
of initial states. Poincar\'e himself seems to corroborate this view as, according to
Barrow-Green (1997), he claimed that ``stable trajectories would {\em outnumber} the unstable, in direct 
analogy with the irrational and rational numbers'' (p. 87, our emphasis).

The  power of Poincar\'e's theorem stems from its being a very general
global result that does not require detailed knowledge of the motion. Of course, it
does presuppose the long-time {\em existence} of solutions! This can be a
matter of concern  due to the possible existence of initial data leading to 
 ``singularities'', that is, obstructions to the extension of solutions. For instance, 
in the Newtonian gravitational $N$-body system the planets (idealized as point particles) 
can collapse or even disperse to infinity in  finite time. Here again, a 
typicality argument comes to the rescue as one hopes to prove that such ``catastrophes'' are 
rare,~\footnote{This also exemplifies the fact that what is taken as  negligible in some contexts may
not be so in others: such ``singularities'' as   collisions of celestial bodies are obviously of 
considerable astronomical interest. Another example are phase-transition points in equilibrium statistical
mechanics.} in the sense that the set of initial conditions leading to them is negligible 
(a very hard problem  which is open in the general case, see Saari, 2005).~\footnote{We should mention
yet another famous (and much harder)  qualitative result in Hamiltonian mechanics: {\em the KAM theorem} 
also can be seen as a typicality result. Roughly, it says that for a sufficiently
small perturbation of an integrable Hamiltonian system (plus technical hypothesis) the set of initial 
conditions leading to quasi-periodic orbits is a set  whose complement has small Lebesgue measure, which
tends to zero as the perturbation tends to zero (P\"oschel, 2000).}

The analogous problem  in kinetic theory  is fortunately much simpler. Consider the usual  
``billiard-balls'' model of a classical gas as made of hard impenetrable spheres (free flow
plus elastic shocks).  There is an ambiguity here as how to extend the flow past an instant 
where three or more particles collide. However, the initial conditions leading to such situation form 
a  measurable subset of lower dimension in phase-space, being therefore of Lebesgue measure zero 
(a corresponding result holds for initial data leading to infinitely many collisions in finite time). Therefore
the dynamics is well defined (for all time) for Lebesgue almost every initial condition (Cercignani, 
Gerasimenko and Petrina, 1997).

As is  well known, the  recurrence theorem was used by Poincar\'e and Zermelo as
a formidable objection to Boltzmann's attempt to reconcile the irreversible character of 
macroscopic phenomena with the reversible nature of the microscopic Newtonian dynamics of 
a gas.~\footnote{Particularly against his ``H-theorem'', see Illner, 1988.} In fact, it
applies to a classical gas of $N$ particles in an isolated bounded container.   But a gas initially 
restricted to  half of the container will, if left to itself,  diffuse until it occupies the whole volume 
while one never observes  it return spontaneously to that initial situation. This is a version of 
the ``recurrrence  paradox'' (or {\em Wiederkehreinwand})  which, together with the ``reversibility objection''
(or {\em Umkehreinwand}), still is  a source of contention among 
physicists.~\footnote{The
author had the opportunity to witness the  extant disagreements  on the occasion of a round table on 
irreversibility at the STATPHYS-20 conference 
in Paris (1998), having professors Ruelle, Lebowitz, Prigogine and Klein as panelists.}

\smallskip

\noindent{\bf 4. Typicality in Statistical Mechanics}

The crucial point in coming to terms with  the reversibility conundrum is the realization that
one is examining a system at two very different {\em levels}. In this sense, Poincar\'e's theorem
is a clue indicating that any hope to derive in a mathematically rigorous way the macroscopic 
irreversible  equations (like Boltzmann's, Euler's or Navier-Stokes) from a  microscopic reversible
dynamics  will involve some kind of idealized limit in which the number of particles goes to infinity
and under which  there is  a ``loss of Poincar\'e-reversibility''.

That some limit procedure is necessarily involved here was clearly stated in
Hilbert's formulation of the 6th problem which, while recognizing  Boltzmann's 
intuition, reads:
\begin{quote}
Thus Boltzmann's work on the principles of mechanics suggests the problem of developing
mathematically the limiting processes, there merely indicated, which lead from the atomistic
view to the laws of motion of continua (cited in Wightman, 1976, p 148).
\end{quote}

A related observation is that under such  change of levels of description there is a
dramatic ``decimation'' of degrees of freedom: while the micro-state has $6N$ (of the
order of $10^{23}$ for a gas in normal conditions) degrees
of freedom (all the particle's positions and velocities) the macro-state usually involves
few  variables (say,  density, pressure and temperature).~\footnote{However, in non-equilibrium
cases one deals with corresponding fields, which are strictly speaking, infinite dimensional
vectors.}

This reduction suggests that some kind of averaging procedure should be involved and which,
together with  the large $N$ limit, points to the  role of statistics. However, 
as there is no intrinsic randomness in the system at hand the use of statistics and probability
might be better understood through typicality. In particular, the taking of averages does not 
necessarily implies any randomness: it  could just mean that  details are not important (Bunge, 1988) 
which is, after all, the basic philosophy behind the qualitative approach. Once more, it  highlights 
\begin{quote}
...what the statistical aspect of statistical mechanics really is, namely, the assertion
of {\em circumstances which may be neglected} (Truesdell, 1966, p.77).
\end{quote}

It then seems that the  general procedure needed to
realize Boltzmann's program in the lines envisioned by Hilbert goes
as follows  (Guerra, 1993, Lanford, 1976). Consider a  ``physically significant'' macroscopic state-function
$\sf F$ (usually linked to locally conserved quantities). A fundamental insight of Boltzmann is that there 
are many different micro-states $\omega \in  \Omega_{\Lambda,N}$ compatible
with the same macrostate  $\sf F$ in the following sense. One partitions the
one-particle phase-space into macroscopically small but microscopically large cells $\Delta_{\alpha}$ and
specifies the number $n_{\alpha}$ of particles lying in each cell when the system is in macro-state $\sf F$ 
(and maybe with additional specifications, like energy, etc, according to desired 
macroscopic description of the system, be it  kinetic, hydrodynamic, etc ). The micro-states corresponding to 
those specifications will have similar density/velocity profiles and define the 
set $\Gamma_{\sf F} \subset \Omega_{\Lambda,N}$.

Let now ${\sf F}(t)$ be the 
macro-state at time $t$ evolved from $\sf F$ according to the
macroscopic phenomenological equations (kinetic, hydrodynamic, etc)
while $\omega(t)$ is the micro-state at time $t$ evolved according to 
Hamilton's equations from an $\omega \in  \Gamma_{\sf F}$. Then,
one would like to prove that for the {\em overwhelming majority} of initial
micro-states $\omega \in  \Gamma_{\sf F}$ and {\em in an appropriate limit} (kinetic,
hydrodynamical, etc) one obtains  $\omega(t) \in  \Gamma_{{\sf F}(t)}$. 
It turns out that  this is  very hard to do  in ``realistic''~\footnote{That is, taking
classical mechanics as the microscopic model. The analogous problem is much more
complete for  stochastic lattice systems, see Spohn, 1991 and Boldrighini, 1996.}
scenarios. In the following we restrict our discussion to a case in which a spectacular 
breakthrough was achieved namely,  Lanford's theorem on the {\em validity problem} for 
Boltzmann's equation.

Recall that Boltzmann (building on previous work of Maxwell) wrote down his 
equation in 1872 to describe the time evolution of the single-particle distribution 
function $f_t({\mathbf r}, {\mathbf v})=f({\mathbf r}, {\mathbf v},t)$ for a dilute gas 
of $N$ identical hard impenetrable spheres of mass $m$ and diameter $a$ in a region $\Lambda \subset \mathbb R^3$. 
It is a  non-linear integral-differential
equation which reads (with no external force field)
\begin{multline}
\frac{\partial}{\partial t} f_t({\mathbf r}, {\mathbf v})+ 
{\mathbf v} \cdot \nabla_{\mathbf r} f_t({\mathbf r}, {\mathbf v})= Q(f_t,f_t)=\\
N a^2 \int_{\mathbb R^3} d^ 3{\mathbf v}_1 
\int_{\mathbf{\hat{n}\cdot}({\mathbf v}-{\mathbf v}_1)\geq0} d{\mathbf{\hat{n}}}\,
{\mathbf{\hat{n}\cdot}}({\mathbf v}-{\mathbf v}_1)
[f_t({\mathbf r}, {{\mathbf v}_1}') \, f_t({\mathbf r}, {\mathbf v}') 
-f_t({\mathbf r}, {\mathbf v}_1) \, f_t({\mathbf r}, {\mathbf v})]
\end{multline}
where  $({\mathbf v},{\mathbf v}_1)$ and  $({\mathbf v}', {{\mathbf v}_1}')$ are, respectively,
the incoming and outgoing velocities; and the so-called  collision operator $Q(\cdot,\cdot)$, 
summarizes the effects of the shocks.

This is one of the most successful equations of physics, with a broad range of applications
and a challenging subject for mathematicians. In particular, and despite some
remarkable recent advances, the global existence and uniqueness of well-behaved 
solutions is still an open problem. However, Boltzmann  ``derived''  his equation by a 
straightforward heuristic analysis  of the collision process using some bold simplifying 
assumptions. Notably, he only considered binary uncorrelated collisions  
(no two particles collide more than once),~\footnote{Ternary and higher order collisions
are of two kinds: ``genuine'' (i.e., simultaneous) which, as we have seen,  are negligible; 
and ``correlated successive binary collisions'' carrying memory effects, which are crucial in the 
study of {\em dense} fluids, see Cohen, 1993.} which boils down to the  famous ``molecular chaos hypothesis''
or {\em Stosszahlansatz}.

But, how can a {\em discrete} $N$-particle classical gas  be described by a {\em continuous}
one-particle distribution function? Sometimes one reads  that 
$N \, f({\mathbf r}, {\mathbf v},t) \, d^3 {\mathbf v} \, d^3 {\mathbf r}$ is
the number of particles of the gas in the infinitesimal region  
$d^3 {\mathbf v} \, d^3 {\mathbf r}$  around  the one-particle phase-space  point  
$({\mathbf v},  {\mathbf r})$ at time $t$. But that cannot be: when the gas is
in the micro-state $\omega(t)={\bf T}_t(\omega(0))=({\bf q}_1(t),{\bf v}_1(t), \ldots, 
{\bf q}_N(t),{\bf v}_N(t))$, the number of particles in, say, a rectangular 
parallelepiped $\Delta \subset \Lambda \times \mathbb R^3$, is given by
$\displaystyle \sum_{i=1}^N {\mathbb I}_{\Delta}({\bf q}_i(t),{\bf v}_i(t))$, which is an
integer.

So, instead of introducing a random ingredient in the system, usually justified 
on the basis of  ``ignorance'' or ``imprecision'' on the initial data, one could take
the viewpoint that $f_t(\cdot, \cdot)$ gives a macroscopic description of the gas as a 
continuum medium,
from which one gets, for instance, the hydrodynamic fields of mass, momentum and kinetic
energy densities respectively:  $\displaystyle \rho({\mathbf r},t)=m N \int_{\mathbb R^3} f_t d^3 {\mathbf v}$, $\displaystyle \rho {\mathbf u}({\mathbf r},t)=m N \int_{\mathbb R^3} {\mathbf v}  f_t 
d^3{\mathbf v}$ and $\displaystyle e({\mathbf r},t) =m N \int_{\mathbb R^3} \frac{1}{2} 
{\mathbf v}^2  f_t d^3{\mathbf v}$.~\footnote{Incidentally, the derivation of hydrodynamic 
(Euler, Navier-Stokes) equations from Boltzmann's equation has a long tradition going back to Hilbert 
and is nowadays quite complete (see Esposito,  Lebowitz and Marra, 1999).
The much harder problem of deriving them from Newton's equations, pioneered by Morrey's work
in the fifties,  is still essentially open.  Recently, Olla, Yau and Varadhan (1993)
were able to get it but at the cost of introducing some unwarranted assumptions (of technical
nature), such as adding  a  small stochastic noise to Newton's equations.}

In other words,   $f$ gives an  approximate (``coarse-grained'' or ``reduced'') 
description of the system which in Lanford's analysis  is made precise as follows (Lanford, 1983): 
a micro-state 
$\omega=({\bf q}_1,{\bf v}_1, \ldots, {\bf q}_N,{\bf v}_N)$ is said to be ``close'' 
to  $f({\bf r}, {\bf v})$
when
\begin{equation}
F_{\Delta}(\omega)= \frac{1}{N} \sum_{i=1}^N {\mathbb I}_{\Delta}({\bf q}_i,{\bf v}_i) \approx
\int_{\Delta} f({\bf r}, {\bf v}) d^3{\bf r} \, d^3{\bf v},
\end{equation}
which  approximation  becomes exact only in an appropriate limit. It was H. Grad who suggested
that the  limit involved here should reflect (in idealized form) the physical situation of
a dilute gas where the diameter of the particles is  much smaller than the mean free path, so
that particles rarely meet. This translates to $N  \rightarrow \infty$ and  $a\rightarrow 0$ 
with $N a^2$ converging to a fixed non-zero constant, called the {\em kinetic limit} or, as suggested 
by Lanford (1976),  {\em Boltzmann-Grad limit} (p. 79). Note the total volume occupied by the particles
is of order $N  a^3$ which goes to zero, such that one is dealing with an ``infinitely diluted gas''.

Now, as one cannot expect the approximation condition to hold for {\em all} micro-states, one 
resorts to a typicality argument to at least guarantee that it will hold for the
 ``vast majority'' of them. Introduce then the following notion: a sequence
$\{{\mathbf P}_{N}\}_{N \geq 1}$ of probability measures on  phase-space~\footnote{Each ${\mathbf P}_{N}$
is defined on the corresponding  phase-space $\Omega_{\Lambda,N}$ which should also be modified to exclude  
overlapping of the  particles.}
is  an  {\em approximating sequence for} $f$ if, for all $\epsilon >0$,
\begin{equation}
{\mathbf P}_N\big[\omega \in  \Omega_{\Lambda,N}: |F_{\Delta}(\omega) - \int_{\Delta} f({\bf r}, {\bf v}) d^3{\bf r} \, d^3{\bf v}|> \epsilon\big] \rightarrow 0,
\end{equation}
in the Boltzmann-Grad limit. This renders precise the micro to macroscopic change of 
description.~\footnote{The concept of approximating sequence also has built-in the
molecular chaos property, see Cercignani, Illner and Pulvirenti, 1994, p. 92.}

We can now state Lanford's theorem, in a very simplified form, as follows:
\begin{quote}
Let  $f_t({\bf r}, {\bf v})$ be a mild solution of Boltzmann's equation with initial
data  $f_0({\bf r}, {\bf v})$. Under some technical hypotheses, if
$\{{\mathbf P}_{N}\}_{N \geq 1}$ is an approximating sequence for $f_0$, then
there exits a $t_0>0$ such that the time-evolved (under the hard sphere dynamics)  
sequence  $\{{\mathbf P}_{N}\circ {\mathbf T}_{-t} \}_{N \geq 1}$ 
is  approximating for  $f_t({\bf r}, {\bf v})$, for all $t \in [0,t_0]$.
\end{quote}

 It is important to realize that Lanford's theorem does not say that Boltzmann's equation
holds ``on the average''  but that it describes the macroscopic behavior of the gas (in the
Boltzmann-Grad limit) for the ``vast majority'' of micro-states  which are 
initially close to $f_0$, at least for some small time interval. The result can
be seen as a {\em law of large numbers} as the conclusion is  that for all $\epsilon >0$
and in the Boltzmann-Grad limit, for $t \in [0,t_0]$:
\begin{equation}
{\mathbf P}_N\big[\omega \in \Omega_{\Lambda,N}: |F_{\Delta}({\bf T}_t(\omega)) - \int_{\Delta} f_t({\bf r}, {\bf v}) d^3{\bf r} \, d^3{\bf v}|> \epsilon\big] \rightarrow 0.
\end{equation}

The rather technical proof, based on a careful analysis of  Liouville's equation and 
the BBGKY-hierarchy (see Cercignani, Illner and Pulvirenti, 1994 or Spohn, 1991),  
requires a ``proper balance between dynamics and probability'' 
(Grad, 1949). As it  turns out,  Lanford's proof is a local result, that is, valid only for a
very short (but strictly positive) time-interval, of the order of one-fifth of a mean free 
path. Though this is a severe shortcoming for applications, it does not diminish the great 
conceptual impact of the result. As remarked by Illner (1988)  ``the limiting evolution given 
by the Boltzmann equation is  irreversible even on such a small time interval'' (p.158). Of course 
a major open problem is to improve the time scale of the  theorem.~\footnote{Results in this direction 
were obtained for a rarefied gas in all space, under  additional hypothesis on the initial data of 
Boltzmann's equation (see Illner and Neunzert, 1989).}  As for the more realistic case of dense fluids 
the situation is much harder (for a discussion, see Cohen, 1997).

Lanford's theorem  is thus the  first and ``remains the only rigorous result
on the scaling limits of many-body Hamiltonian systems with no unproven assumptions'' (Yau,
1998, p.194). It can be seen, even with all its restrictions, as the realization, after nearly a 
hundred years, of Boltzmann's intuitions as made precise by Hilbert, Grad, Morrey and many others.
According to Gallavotti (1999),
 ``this is an important confirmation, mathematically rigorous, of
Boltzmann's point of view according to which reversibility, and the corresponding recurrence
times, is not in contradiction with the experimental observation of irreversibility'' (p. 35).

\smallskip

\noindent{\bf 4. Conclusions}

The standard textbook justification for the use of probability 
in classical statistical mechanics  follows an operationalistic view. First, as the microscopic
dynamics itself is non-random, any randomness is shifted to the initial conditions. Then, goes
the argument, due to our inability to either solve the huge number of
equations or  measure with  precision the initial data,
we have to resort to statistics. In other words, human limitations
are the basis for the justification.

We find that untenable and argued in favor of
an alternative viewpoint based on the notion of typicality.  
This is not to say that other viewpoints would not be more adequate in other 
contexts,  but that in the case of {\em classical} statistical mechanics typicality seems to be more natural.
It does not invoke any randomness (ontological or epistemological), which is  consistent 
with the  kind of classical  system at hand. Also, as we have seen, measure-theoretical typicality 
arguments have been used successfully in the qualitative study of Hamiltonian systems in many other 
contexts. The idea is to obtain  results valid for the  ``overwhelming majority'' of initial 
conditions. In this sense, as the initial conditions are an integral part of mechanical systems, 
probability-as-typicality is not that foreign to mechanics.

Of course, one {\em does} need an ingredient ``outside of mechanics'' when trying to bridge
the micro and macro levels of description which is the main goal of statistical mechanics.
Boltzmann had the intuition that some idealized limit would be involved  and, as 
Lanford's analysis illustrates,  in classical statistical mechanics probability measures enter as
crucial level-connecting concepts in realizing that goal in a rigorous way.  Also, Lanford's 
theorem does imply that for some ``rare''  initial conditions the corresponding macroscopic dynamics
will not follow the observed behavior. However, because they are rare in the measure-theoretic
sense used in statistical mechanics, the idea is that such  data can be ignored. This is in the spirit of 
measure/probability theory in which a property is taken to hold true 
when the set of exceptions is rare in the sense of having  very small measure/probability (even if  such
set is large in terms of cardinality). In this sense, we
suggest that probability-as-typicality is a way to express, in a mathematically  precise  (albeit idealized) 
way, the validity of the macroscopic  laws and their  compatibility with the atomistic-mechanical  microscopic
model.  

As usual in mathematical-physics, the idealizations are the
unavoidable price to pay in exchange for  rigorous analysis. Another example: 
in equilibrium statistical mechanics in order to define  phase-transitions points as
singularities of the partition functions one has to take the thermodynamic limit. This does not 
mean that real (finite) physical systems do not exhibit phase-transitions but that 
the idealization expressed by the limit  $N \rightarrow  \infty$ helps in better
handling the problem mathematically  than with a finite system.~\footnote{As remarked by the late
mathematical-physicist  R. Dobrushin (1997),  ``infinity is a better approximation to
the number $10^{23}$ than the number 100'' (p.227).}

Many tough questions are still to  be addressed. For example, as typicality is relative
to the measure used, how one  justifies a particular choice? Under what criteria? Moreover, what about
other notions of typicality? Also, as we mentioned, typicality is not enough
to decide when a given initial data belong to a desired subset, which is very
important regarding the trend to equilibrium issue. And of course, typicality  does not avoid the need of 
rigorous  analysis and proof.

Finally, though  the notion of probability-as-typicality is not new, it is seldom articulated 
clearly and  it should be allowed more space in the debates on 
the foundations of statistical mechanics. We also think that the appearance of that notion at 
approximately the same time in the theory of dynamical systems and statistical mechanics  deserves a 
deeper investigation as it could be seen as a symptom  of that broad historical
transition  from  quantitative to the qualitative methods.

\bigskip
 
\noindent{\bf ACKNOWLEDGMENTS} 
The author would like to thank the anonymous referees for their criticisms and suggestions
which helped improving the text.

Work partially supported by
FINEP (Pronex, ``Fen\^omenos Cr\'iticos em Probabilidade e Processos Estoc\'asticos'') and FAPERJ.
\bigskip

\noindent{REFERENCES}

\vspace{-2mm}
\noindent\hrulefill


\begin{trivlist}

\smallskip

\smallskip
\item Barrow-Green, J. (1997). {\em Poincar\'e and the three-body problem}. History
of Mathematics, vol. 11, American Mathematical Society and London Mathematical Society.

\smallskip
\item Boldrighini, C. (1996). Macroscopic limits of microscopic systems. {\em Rediconti
di Matematica}, Serie VII, Vol. 16, 1-107.

\smallskip
\item Brush, S. G. (1994). {\em The kind of motion we call heat. A history of the kinetic
theory of gases in the 19th century. 1. Physics and the Atomists}. New York: North Holland.

\smallskip
\item Bunge, M. (1967). {\em Foundations of Physics}. New York: Springer-Verlag.

\smallskip
\item  Bunge, M. (1988). Two faces and three masks of
probability. In  E. Agazzi (ed.), {\em Probability in the Sciences} (pp.27-50). New York:
Kluwer Academic Publishers.

\smallskip
\item Cercignani, C. (1998). {\em The man who trusted atoms}. Oxford: Oxford University
Press.

\smallskip
\item Cercignani, C., Illner, R. and Pulvirenti (1994). {\em The mathematical
theory of dilute gases}. New York: Springer-Verlag.

\smallskip
\item Cercignani, C., Gerasimenko, V.I. and Petrina, D. Ya. (1997). {\em Many-particle
dynamics and kinetic equations}. Boston: Kluwer Academic Publishers..

\smallskip
\item Chenciner, A. (1999). De la M\'ecanique C\'eleste \`a la th\'eorie des Syst\`emes 
Dynamiques, aller et retour.\\
{\texttt www.imcce.fr/Equipes/ASD/person/chenciner/chen\_preprint.php}

\smallskip
\item  Choquet, G. (2004). Borel, Baire, Lebesgue. In Choquet, G., De Pauw, T., de la Harpe, P.,
Kahane, J.-P., Pajot, H. and S\'evennec, B. (Eds.), {\em Autour du centennaire
Lebesgue- Panoramas \& Synth\'eses}, 18 (pp. 23-37). Marseille: Soci\'et\'e Math\'ematique de France.

\smallskip
\item  Cohen, E.G.D. (1993). Kinetic theory: understanding nature through collisions,
{\em American  Journal of  Physiscs}, 61(6), 524-533. 

\smallskip 
\item Cohen, E. G. D. (1997). Bogolubov and kinetic theory: the Bogolubov equations.,
{\em Mathematical models and methods in applied sciences}, vol. 7, no. 7, 909-933.

\smallskip
\item   Corry, L. (1997). David Hilbert and the axiomatization of
physics, {\em Archive for  History of Exact Sciences}, 51,  83-198.

\smallskip
\item  Diaconis, P., Holmes, S.  and Montgomery, R. (2004). 
Dynamical Bias in the 
Coin Toss.
{\texttt www.count.ucsc.edu/~rmont/papers/headswithJ.pdf}

\smallskip
\item Dobrushin, R. L. (1997). A mathematical approach to foundations of
statistical mechanics. In Cercignani, C., Jona-Lasinio, G., Parisi, G. and Radicati di Brozolo, L. A.
{\em Boltzmann's legacy 150 years after his birth},
Atti dei Convegni Lincei 131 (pp.227-243). Roma: Academia Nacionale dei Lincei.

\smallskip
\item D\"urr. D. (2001). Bohmian Mechanics.   In
Bricmont, J., D\"urr, D., Gallavotti, M.C., Ghirardi, G., Petruccione, F. and
Zangh\`i (Eds.), {\em Chance in Physics} (pp. 115-131). New York: Springer.

\smallskip
\item Esposito, R., Lebowitz, J. L. and Marra, R. (1999). On the Derivation of Hydrodynamics from the 
Boltzmann Equation. {\em Physics of Fluids}, 11, 2354-2366.

\smallskip 
\item Gallavotti, G. (1999). {\em Statistical Mechanics: A Short Treatise}.
New York: Springer.

\smallskip
\item Goldstein, S. (2001). Boltzmann's approach to statistical mechanics. In
Bricmont, J., D\"urr, D., Gallavotti, M.C., Ghirardi, G., Petruccione, F. and
Zangh\`i (eds.), {\em Chance in Physics} (pp.39-54). New York: Springer.

\smallskip 
\item Goroff, D. L. (1992). Henri Poincar\'e and the birth of chaos theory: an
introduction to the english translation of {\em Les M\'ethodes Nouvelles de la
M\'ecanique C\'eleste}. In  Poincar\'e, H. and Goroff, D. L. (Ed.), {\em
New Methods of Celestial Mechanics (History of Modern Physics)}, Vol. 1 (pp. I1-I107). AIP Press.

\smallskip
\item Grad, H. (1967). Levels of description in statistical mechanics and
thermodynamics. In M. Bunge (Ed.), {\em Delaware seminar on foundations of physics},
Studies in Foundations, Methodology, vol. 1 (pp.46-79). Berlin-Heidelberg-New York: Springer-Verlag.

\smallskip
\item Guerra, F. (1993). Reversibilidade/Irreversibilidade. In R. Romano (Ed.),
{\em Enciclop\'edia Einaudi}, no. 24. Porto: Imprensa Nacional-Casa da Moeda.

\smallskip
\item  Henkel, M. (2001). Sur la solution de Sundman du probl\`eme des trois corps, 
{\em Philosophia Scientiae}, 5 (2), 161-184. 

\smallskip
\item Hoare, G. T. Q. and Lord, N. J. (2002). `Int\'egrale, longueur, aire' the centenary
of the Lebesgue integral. {\em The mathematical gazette}, march, 3-27.

\smallskip 
\item Illner, R. and Neunzert, H. (1987). The concept of reversibility in the
kinetic theory of gases. {\em Transport theory and statistical physics}, 16 (1),
89-112.

\smallskip
\item Illner, R. (1988). Derivation and validity of the Boltzmann equation: some
remarks in reversibility concepts, the H-functional and coarse-graining. In  J. M. Ball (Ed.), {\em Material
instabilities in continuum mechanics and related mathematical problems}  (pp.147-174). Oxford:
Clarendon Press.

\smallskip 
\item Illner, R. and Neunzert, H. (1989). Global validity of the Boltzmann equation for
a two- and three-dimensional rare gases  in vacuum: Erratum and improved results. {\em Communications
in Mathematical  Physics}, 121, 143-146.

\smallskip
\item Kac, M. (1949). {\em Statistical independence in probability, analysis and
number theory}, The Carus Mathematical Monographs, no. 12, MAA.

\smallskip
\item  Kahane, J.-P. (2001). Naissance et posterit\'e de l'int\'egrale de Lebesgue.,
{\em Gazette des Math\'ematiciens}, 89, 5-20. SMF.

\smallskip
\item  Kolmogorov, A. N. (1956).  {\em Foundations of the
Theory of Probability}. New York: Chelsea. 

\smallskip
\item Laskar, J. (1992). La stabilit\'e du syst\`eme solaire. In Dalmedico, A.D.,
Chabert, J.-L. and Chemla, K. (Eds.), {\em Chaos
et d\'eterminisme} (pp. 171-211). Paris: \'Editions du Seuil.

\smallskip
\item Lanford, O. E. (1983). On a derivation of the Boltzmann equation. In J. L. Lebowitz and E. W.
Montroll (Eds.), {\em Non-equilibrium Phenomena I} (pp. 3-17). New York:  North-Holland.

\item Lanford, O. E. (1976). Time evolution of  large classical systems. In J. Moser (Ed.),
{\em Lecture notes in physics}, 38 (pp. 1-111). Berlin: Springer.

\smallskip
\item  Mazliak, L., Chaumont, L. and
Yor, M. (2004). Kolmogorov: Quelques aspects de l'oeuvre probabiliste. In {\em
L'h\'eritage de Kolmogorov en math\'ematiques}, Editions Belin. 

\smallskip
\item Olla, S., Varadhan, S. R. S. and Yau, H. T. (1993). Hydrodynamical limit
for a Hamiltonian system with weak noise. {\em  Communications in  Mathematical Physics}, 61, 119-148.

\smallskip
\item Oxtoby, J. L. (1971). {\em Measure and Category}. New York  Springer.

\smallskip
\item P\"oschel, J. (2000). A lecture on the classical KAM theorem. {\em Proceedings of  Symposia
in Pure Mathematics}, 69, 707-732.

\smallskip
\item Saari, D. G. (2005). {\em Collisions, Rings and Other Newtonian
N-Body Problems}, CBMS, Regional Conference Series in Mathematics, Number 104. Rhode Island: American Mathematical Society.

\smallskip
\item Spohn, H. (1991). {\em Large scale dynamics of interacting particles}. Berlin: Springer.

\smallskip
\item Torretti, R. (1999). {\em The philosophy of physics}. Cambridge: Cambridge University Press.

\smallskip
\item Truesdell, C. A. (1966).  The ergodic problem in classical
statistical mechanics. In {\em Six Lectures on Modern Natural Philosophy}. New York: Springer.

\smallskip
\item Uffink, J. (2004). Boltzmann's work in statistical physics. In {\em Stanford
Encyclopedia of Philosophy}.

\smallskip
\item  Volchan, S. B. (2002). What is a random sequence? {\em American  Mathematical Monthly}, 109,  No.1, 
January, 46-63.

\smallskip
\item von Plato, J. (1983). The method of arbitrary functions. {\em British  Journal of  Philosophy
of Science}, 34, 37-47.

\smallskip 
\item von Plato, J. (1991). Boltzmann's Ergodic Hypothesis. {\em Archive for History of Exact Sciences}, 42,
71-89.

\smallskip
\item  von Plato, J. (1998) {\em Creating Modern Probability}. Cambridge:
Cambridge University Press.

\smallskip
\item Yau, H.-T. (1998). Scaling Limit of Particle Systems, Incompressible Navier-Stokes Equation and Boltzmann Equation. {\em Documenta Mathematica Journal}, DMV, Extra Volume ICM III, 193-202.

\smallskip
\item Yoccoz, J.-C., (1995). Recent developments in dynamics. In {\em Proceedings of the International Congress of Mathematicians}, {\bf 1} (pp. 247-265). Basel: Birkh\"auser.

\smallskip
\item Wick, D. (1995). {\em The infamous boundary}. New York: Springer.

\smallskip
\item Wightman, A. S. (1976). Hilbert's sixth problem: mathematical
treatment of the axioms of physics. In {\em Proceedings of Symposia
in Pure Mathematics}, vol. 28, AMS, 147-240

\smallskip

\end{trivlist}

\bigskip

\noindent{\bf S\'ERGIO B. VOLCHAN} 

\noindent{\em
Departamento de Matem\'atica, Pontif\'{\i}cia Universidade Cat\'olica do 
Rio de Janeiro, Rua Marqu\^es de S\~ao Vicente 225,
G\'avea, 22453-900  Rio de Janeiro, Brasil

\noindent volchan@mat.puc-rio.br}
\end{document}